\documentclass[apj]{emulateapj}

\shorttitle{Stellar structure of the Milky Way}
\shortauthors{de Jong et al.}

\begin{document}

\title{Mapping the stellar structure of the Milky Way thick disk and halo
  using SEGUE photometry}

\author{Jelte T. A. de Jong\altaffilmark{1},
Brian Yanny\altaffilmark{2}, Hans-Walter Rix\altaffilmark{1},
Andrew E. Dolphin\altaffilmark{3}, Nicolas F. Martin\altaffilmark{1},
Timothy C. Beers\altaffilmark{4} }

\email{dejong@mpia.de}

\altaffiltext{1}{Max-Planck-Institut f\"ur Astronomie, K\"onigstuhl
  17, D-69117 Heidelberg, Germany}
\altaffiltext{2}{Fermi National Accelerator Laboratory, P.O. Box 500,
  Batavia, IL 60510, United States} 
\altaffiltext{3}{Raytheon Company, 1151 E Hermans Rd, Tucson, AZ
  85756, United States}
\altaffiltext{4}{Department of Physics \& Astronomy and JINA: Joint
  Institute for Nuclear Astrophysics, Michigan State University,
  E. Lansing, MI, 48824, USA}

\begin{abstract}
  We map the stellar structure of the Galactic thick disk and halo by
  applying color-magnitude diagram (CMD) fitting to photometric data
  from the SEGUE survey.  The SEGUE imaging scans allow, for the first
  time, a comprehensive analysis of Milky Way structure at both high
  and low latitudes using uniform SDSS photometry.  Incorporating
  photometry of all relevant stars simultaneously, CMD fitting
  bypasses the need to choose single tracer populations.  Using old
  stellar populations of differing metallicities as templates we
  obtain a sparse three-dimensional map of the stellar mass
  distribution at $|Z|>$1 kpc.  Fitting a smooth Milky Way model
  comprising exponential thin and thick disks and an axisymmetric
  power-law halo allows us to constrain the structural parameters of
  the thick disk and halo. The thick-disk scale height and length are
  well constrained at $0.75\pm0.07$ kpc and $4.1\pm0.4$ kpc,
  respectively. We find a stellar halo flattening within $\sim$25 kpc
  of $c/a=0.88\pm0.03$ and a power-law index of $2.75\pm0.07$ (for 7
  kpc $\lesssim R_{GC} \lesssim$~30 kpc). The model fits yield
  thick-disk and stellar halo densities at the solar location of
  $\rho_{thick,\sun}=10^{-2.3\pm0.1}$ $M_\sun$pc$^{-3}$ and
  $\rho_{halo,\sun}=10^{-4.20\pm0.05}$ $M_\sun$pc$^{-3}$, averaging
  over any substructures. Our analysis provides the first clear in
  situ evidence for a radial metallicity gradient in the Milky Way's
  stellar halo: within $R\lesssim$15 kpc the stellar halo has a mean
  metallicity of $[Fe/H]\simeq-1.6$, which shifts to
  $[Fe/H]\simeq-2.2$ at larger radii, in line with the two-component
  halo deduced by \cite{carollo07} from a local kinematic
  analysis. Subtraction of the best-fit smooth and symmetric model
  from the overall density maps reveals a wealth of substructures at
  all latitudes, some attributable to known streams and overdensities,
  and some new. A simple warp cannot account for the low latitude
  substructure, as overdensities occur simultaneously above and below
  the Galactic plane.
\end{abstract}


\section{Introduction}
\label{sec:intro}

Studying the stellar structure of the Milky Way has a long history, as
early models based on star counts were already constructed by the
likes of \cite{herschel1785} and \cite{kapteyn1922}. Today we know
that four `components' make for a sensible approximate description of
the Milky Way stellar body, each with different structural, kinematic
and population characteristics: the bulge, the thin and the thick
disk, and the stellar halo. Accurate determination of the properties
of these components is a difficult undertaking, as it requires surveys
with (1) sufficient sky coverage to assess the overall geometry, (2)
sufficient depth for mapping stars to $\sim$10 to 30 kpc, and (3)
sufficient information (e.g., color-based luminosity estimates) to
obtain reasonable distance estimates for these stars. Furthermore, the
presence of dust in the plane of the disk clouds our view of most of
the Galaxy at wavelengths shorter than a few microns. For this reason,
most previous efforts to study Milky Way structure at low latitudes
have used (near-)infrared data, such as the work by \cite{kent91}
based on data from the Spacelab infrared telescope, or more recent
work using 2MASS data \citep[e.g.][]{momany06,reyle09} or the GLIMPSE
survey \citep{benjamin05}. While near-infrared surveys provide the
only way of studying the low latitude regions, because they suffer
much less from dust extinction, they are currently less sensitive than
their optical counterparts and often rely on tracer populations with
less accurate distances.

Over the past few years, data from 2MASS \citep{2mass} and the Sloan
Digital Sky Survey \citep[SDSS;][]{york00,dr7} have played a key role
in revolutionizing our empirical picture of the stellar components of
the Milky Way. As the SDSS survey avoided the Galactic plane, much of
the work using SDSS data has focused away from the bulk of the Milky
Way's stars to high galactic latitudes. Nonetheless, \cite{juric08}
have been able to constrain the structural parameters of the stellar
halo and the thin and thick disk.  Beyond the global parameters, these
surveys have also led to the discovery of substructure in both the
stellar disk \citep{newberg02,ibata03,martin04,bellazzini06} and halo
\citep{newberg02,fieldofstreams,grillmair06,hercaqui,juric08,bell08}.
This detailed substructure provides information on how the Milky Way
formed and evolved; as galaxies assemble through mergers and
accretions of smaller systems, these events leave identifiable
signatures in their structure and kinematics.  This also has
cosmological implications, as the structure of the dark halo and the
interactions with dark subhalos can significantly influence the
appearance of a galaxy \citep[e.g.][]{kazantzidis08,younger08}.

In this paper we derive a new map of the stellar distribution in the
Milky Way, in order to study the large-scale structure of the thick
disk and stellar halo. Though covering only the Northern celestial
hemisphere, and that only sparsely, this map covers both high and low
latitudes and uses rigorously derived stellar distances. This map is
based on one of the constituent projects of the extended SDSS survey
(SDSS II), SEGUE \citep[Sloan Extension for Galactic Understanding and
Exploration; ][]{segue}, which is an imaging and spectroscopic survey
aimed at the study of the Milky Way and its stellar
populations. Whereas the main SDSS survey avoided low Galactic
latitudes, SEGUE imaging scans go through the Galactic plane (see
Figure \ref{fig:dataoverview}), allowing a `picket-fence' view of the
Galaxy at these low latitudes. It is this deep, uniform data set that
enables a systematic analysis of stellar structure as function of
Galactic latitude and longitude, which we base on color-magnitude
diagram (CMD) fitting techniques developed by \cite{sdssmatch}. A
crucial advantage of an analysis based on CMDs is that they bypass the
often ad-hoc and sometimes problematic choices of stellar `tracers'
(such as K-dwarfs or F-stars) to delineate the structure. Instead,
this procedure results immediately in a mass-density map for stellar
population components of, say, different metallicity.

To avoid having to fit the complex stellar population make-up of the
thin disk, we in the end consider only stars that are more than 1 kpc
away from the Galactic mid-plane, and treat the thin disk as a small
perturbation to our results. As SEGUE is a northern hemisphere survey,
it stays far away from the Galactic bulge, which can be safely
ignored. On the other hand, the resulting maps of stellar mass density
in the thick disk and halo allow us to measure the structural
parameters of these components. In addition, we show that subtracting
a smooth Galactic model reveals a wealth of substructure, especially
at low latitudes, which we will study in detail in a forthcoming
publication.

The outline of this paper is as follows. Section \ref{sec:data}
describes the data used for our analysis, and \S\ref{sec:cmdfitting}
describes the CMD-fitting techniques we utilize. The resulting stellar
distributions are presented in \S\ref{sec:results}, where we also fit
smooth Galactic models to them, derive structural parameters, and look
at deviations from the best fitting model. Finally, we summarize and
discuss our results in \S\ref{sec:discussion}.

\section{SEGUE photometry and coverage}
\label{sec:data}

\begin{figure}[t]
\epsscale{1.2}
\plotone{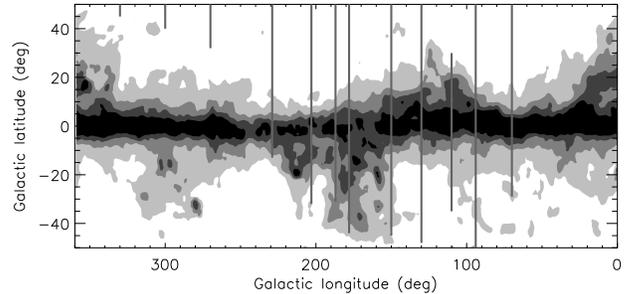}
\caption{ Overview of low-latitude imaging scans from SEGUE
\citep{segue} used for this analysis, shown as the vertical lines in
Galactic coordinates. The grayscale shows dust reddening estimates
based on the extinction maps from \cite{sfd}, with increasingly dark
regions corresponding to $E(B-V) >$0.1, 0.25, 0.5 and 1.0 mag. Only
regions with $E(B-V)\leq$ 0.25 are used in the analysis.}
\label{fig:dataoverview}
\end{figure}

As part of the SEGUE survey \citep{segue}, ten 2.5\degr-wide scans,
crossing the Galactic plane at fixed longitudes, were imaged in five
passbands \cite[$u$, $g$, $r$, $i$, and $z$;][]{Gu98,Gu06,Ho01}. These
data are available as part of SDSS Data Release 7
\citep[DR7,][]{dr7}. An automated data reduction and analysis pipeline
produces accurate astrometric and photometric measurements for all
detected objects \citep{Lu99,St02,Sm02,Pi03,Iv04,Tu06}, reaching a
photometric accuracy of 2\% down to $g\simeq$22.5 and $r\simeq$22
\citep{Iv04,sesar07}. Figure \ref{fig:dataoverview} shows the coverage
of the scans between Galactic latitudes of +50\degr~and $-$50\degr.
Most scans have some overlap with the SDSS main survey or SDSS Legacy
area in the North Galactic cap, and we extract 2.5\degr-wide strips of
data from DR7 to extend the photometry to Galactic latitudes of
85\degr. For the CMD fitting analysis we restrict ourselves to two
bands, $g$ and $r$.  These two bands are the most useful for our
population/distance analysis because, among the three photometrically
most sensitive bands ($g$, $r$ and $i$), they combine to give the
largest offset in magnitude between stellar main sequences of
different metallicities.  The SEGUE imaging in DR7 have been
photometrically calibrated following the procedure of
\cite{padmanabhan08}.  The procedure basically uses the overlap of the
SEGUE data scans with the Legacy SDSS imaging scans to solve for the
photometric zero-points and flat-field co-efficients of the SDSS 2.5m
camera system.  This overlap generally occurs in uncrowded regions of
sky ($|b| > 20$\degr) even for scans which extend to cross the
Galactic plane.  The photometric solution for the SEGUE scans was done
incrementally rather than globally after the global solution for all
the SDSS Legacy scans.  The accuracy of the photometry is estimated to
be to about 1.5\% in g, and r ($<2\%$ in color). See \S 2.2 of
\cite{segue} for more details on the SEGUE imaging and calibration.
We de-redden all data using the dust maps from \cite{sfd}, including
the correction suggested by \cite{bonifacio00}, provided through the
SDSS catalog server. As most of the dust is confined close to the
Galactic plane, it is in the foreground for all stars more distant than
1 kpc from the mid-plane which we employ in our analysis.

\begin{figure}[t]
\epsscale{1.0}
\plotone{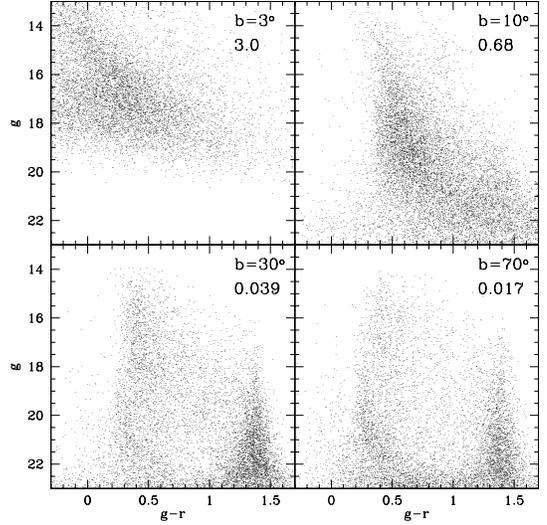}
\caption{ Typical color-magnitude diagrams taken from the SEGUE
imaging scan at $l$=94\degr, dereddened based on the dust extinction
maps from \cite{sfd}. In the upper right corner of each panel the
Galactic latitude and the average reddening in $E(B-V)$ \citep{sfd}
corresponding to each CMD are indicated. The latitude range from which
the CMDs were extracted were chosen so that each contains
approximately the same number of stars.}
\label{fig:examplecmds}
\end{figure}

To illustrate the basis for our subsequent modeling, several
de-reddened CMDs drawn from different latitudes in the SEGUE stripe at
$l$=94\degr are shown in Figure \ref{fig:examplecmds}. The most
obvious difference between different latitudes is the variation in
reddening, which ranges from an $E(B-V)$ of up to several magnitudes
very close to or in the plane of the Galaxy to a few hundredths of a
magnitude at high latitudes. Once the reddening reaches several tenths
of magnitudes (e.g., the upper right panel of Fig. \ref{fig:examplecmds}),
the unaccounted differential reddening along the line of sight smears
out the CMD features. For this reason, and because the dereddening
itself is less reliable in high extinction regions, we exclude regions
with $E(B-V)>$0.25 from our analysis; the excluded regions correspond
to the three darkest shadings in Fig. \ref{fig:dataoverview}. Hence,
the two lower panels in Figure \ref{fig:examplecmds} are
representative for the CMDs used for this study. This cut in
foreground extinction also means that we exclude the most crowded
regions and thereby avoid possible problems due to crowding.

\begin{figure}[t]
\epsscale{1.0}
\plotone{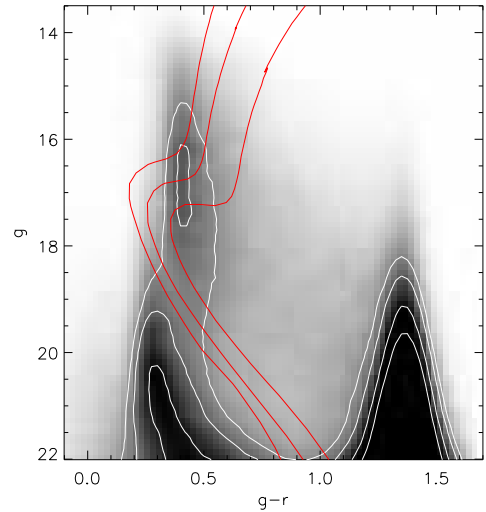}
\caption{ A Hess diagram (greyscale) of the SEGUE scan at $l$=94\degr
and $b>$30\degr, in dereddened ($g$,$g-r$). Overplotted are isochrones
corresponding to the three template stellar populations fit to the
data. From right to left the isochrones are for the `thick-disk-like'
([Fe/H]=$-$0.7), the `inner-halo-like' intermediate metallicity
([Fe/H]=$-$1.3) and the `outer-halo-like' metal-poor ([Fe/H]=$-$2.2)
populations.}
\label{fig:isochrones}
\end{figure}

Before carrying out a more formal CMD analysis, it is instructive to
describe the main features seen in the CMDs and to realize which kind
of stars they contain. Figure \ref{fig:isochrones} shows the number
density of stars in bins in the color-magnitude plane, or Hess
diagram, of all stars in the SEGUE scan at $l$=94\degr~and
$b>$30\degr. The large concentration of stars in the lower right,
i.e. at faint magnitudes and centered at a $g-r$ color of $\sim$1.4
consists of nearby, intrinsically faint and low-mass dwarf stars in
the disk. The other plume of stars at $g-r\sim$0.4 is where the
main-sequence turn-off (MSTO) stars of old ($\gtrsim$10 Gyr) stellar
populations lie.  A shift in the average color of this plume of stars
is visible around $g=$18. This is caused by the difference in
metallicity between the thick disk, which dominates at brighter
magnitudes, and the Galactic halo, which dominates at larger
distances.  The color of the MSTO is a useful proxy for the
metallicity of the dominant population at a given distance
\citep[e.g. see the discussion about this for SDSS isochrones
in][]{girardi04}.  Assuming a regular and smooth Milky Way, the main
change between different latitudes are the relative contributions of
(thick) disk and halo stars to the CMD. For example, comparing the
CMDs corresponding to latitudes of 30\degr~and 70\degr~(lower panels
of Fig. \ref{fig:examplecmds}) shows how the number of disk stars
decreases with respect to the halo stars. The field at
70\degr~latitude adds complexity to this description, as it contains
an example of stellar substructure; the main-sequence feature produced
by the Sagittarius stream is readily visible, running from
($g$-$r$,$g$)$\simeq$(0.2,20.0) to (0.5,22.0).

For our analysis of stellar structure we divide the 2.5\degr-wide
SEGUE stripes into bins of 1\degr width in latitude from which CMDs
are created. Each CMD is then analysed as described in the next
Section.

\begin{figure*}[t]
\epsscale{1.0}
\plotone{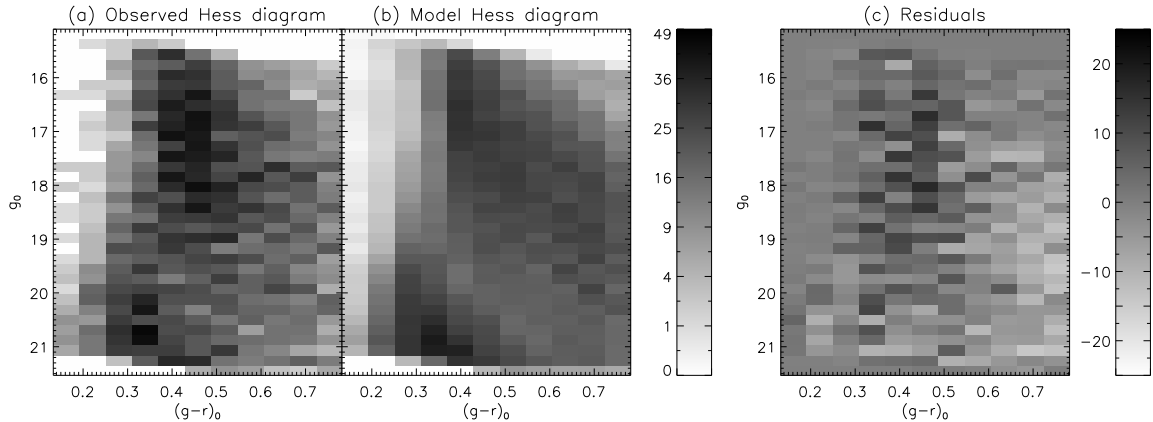}
\caption{ Example of distance distribution fitting, based on the
Hess/CMD diagram at $l$=94\degr~and $b$=+30\degr; note that only the
blue portion of the full color range in Figure \ref{fig:isochrones} is
shown here. No attempt to model low-mass thin disk stars dominating at
$g-r>$0.8 is made here. {\it (a)} Observed distribution of stars in
the color-magnitude plane. {\it (b)} Best-fit model to (a), consisting
of a combination of template stellar populations at different
distances.  {\it (c)} Residuals after subtracting the model (b) from
the data (a). Color bars indicate the number of stars in the Hess
diagram bins. The mean residuals across the CMD are $\sim$25\% of the
mean star counts across the entire CMD.}
\label{fig:examplehess}
\end{figure*}

\section{Distance fitting of CMDs}
\label{sec:cmdfitting}

In most applications of CMD fitting, observed photometry is compared
with models in order to determine which combination of simple stellar
populations best resembles the data, thereby providing a model of the
star formation history (SFH) and age-metallicity relation (AMR). For
the work presented here, we fit model stellar population CMDs based on
the isochrones in SDSS filters provided by \cite{girardi04} to the
SEGUE photometry. The software package we use for this \citep[MATCH,
][]{dolphin97,match} uses a maximum-likelihood method to determine the
best linear combination of models, after transforming both models and
data to Hess diagrams (2D histograms of stars in the color-magnitude
plane), enabling a pixel-by-pixel comparison. For a proper comparison
the models need to be convolved with a realistic photometric error and
completeness model. The model we use here is the same as in
\cite{sdssmatch}, based on the photometric errors from the SDSS
pipeline \citep{Iv04} and the completeness determined from a
comparison\footnote{See
http://www.sdss.org/dr5/products/general/completeness.html} between
SDSS and
COMBO-17\footnote{http://www.mpia.de/COMBO/combo\_index.html}. Since
foreground reddening precludes the use of the most crowded regions in
the mid-plane, crowding does not significantly affect the completeness
in the data used here.

Rather than determining the SFH and AMR, our aim in the present
context is to map the structure of the stellar populations in the
inner and outer Milky Way. For this, we use the distance-fitting
option of MATCH, described and tested in \cite{sdssmatch}. To limit
the number of free parameters and parameter degeneration in the fits,
we define a small set of template stellar populations.  The
choice of templates is motivated both by the populations we expect to
be present in the outer disk and inner halo, and by the CMD features
we have to fit. As discussed in \S \ref{sec:data}, the most obvious
indicator of population differences will be the MSTO color, which
means the templates should span the expected color range, preferably
in regular intervals. The MSTO color of a stellar population depends
both on its age and metallicity, but since both the thick disk and the
halo stellar populations are known to be old, we choose to use a fixed
age range and use metallicity offsets to probe the MSTO color range.
We adopt three different metallicity bins, [Fe/H]=$-$0.7, $-$1.3 and
$-$2.2, respectively. All template populations have the same age range
of 10.1$<$log[t/yrs]$<$10.2, and an assumed binary fraction of
0.5. For convenience, we choose a nomenclature for these components
that is based on previous work: we refer to [Fe/H]=$-$0.7 as a
thick-disk-like population and, following \cite{carollo07}, refer to the
[Fe/H]=$-$1.3 and $-$2.2 templates as the two halo populations,
namely, an inner-halo-like and outer-halo-like population,
respectively. Note, however, that in our subsequent analysis this
choice of terms neither immediately prejudges the geometry, nor
implies that these components are `distinct'. We simply presume that
these three components are sufficient to completely describe the
stellar populations at any point that is at least 1 kpc above or below
the Galactic mid-plane. Isochrones corresponding to these population
templates are overplotted on Figure \ref{fig:isochrones}. The colors
of the MSTO of populations 1 and 2 match the color of the prominent
upper and lower plume of MSTO stars in Figure \ref{fig:isochrones},
respectively, while the third population has a bluer MSTO. Using a
coarse grid of metallicities means that for stars with a metallicity
different from any of the template populations a discrepant distance
is inferred, as the brightness of the MS depends is metallicity
dependent (see Fig. \ref{fig:isochrones}). The magnitude offsets in
this case are $\sim$0.3 and $\sim$0.5 magnitudes, going from the
highest ([Fe/H]=$-$0.7) to the lowest metallicity ([Fe/H]=$-$2.2)
isochrone. As stars with intermediate metallicities will be
interpolated between the template populations, this translates into
distance uncertainties of at most 10\%. The other contributions to the
distance uncertainty are inaccuracies in the isochrones and the
photometric uncertainties, both of which are expected to be much
smaller. The overall distance uncertainty is therefore of order 10\%.

We deliberately choose to avoid fitting the thin disk, because a
template population with more typical thin-disk populations should
have a large range in age and metallicity and would be difficult to
distinguish from a combination of the three templates described
above. Hence, when presenting and interpreting our results, we will
ignore the distance ranges where the thin disk is expected to be an
important contribution to the star counts, and only consider regions
with $Z>$1.0 kpc, or roughly four thin-disk scale heights.

All stars with 15$<g<$21.5, 15$<r<$21.0 and 0.1$<g$-$r<$0.8 are used
in the fits, with Hess diagram pixel sizes of 0.07 in color and 0.2 in
magnitude. This resolution is sufficient to resolve the features on
which the analysis depends, but still ensures sufficient
signal-to-noise per pixel. The color-cut of $g$-$r<$0.8 eliminates the
issue of fitting the faint, red thin-disk stars. For each of the
templates, model CMDs are created for a range of distance moduli. The
magnitude limits used in the fits (15$<g<$21.5) correspond to distance
limit for MSTO stars of roughly 1.5$<D<$25 kpc. Since the MSTO stars
provide the best constraints for these CMD fits, only this distance
range will be analysed. However, to avoid edge effects, model
templates are created for distance moduli between 5.0 (100 pc) to 20.0
(100 kpc) in steps of 0.2 mag ($\sim$10\% distance bins).  MATCH then
determines the best-fitting combination of model CMDs. Subsequently,
the uncertainties on the obtained results are determined by refitting
Monte Carlo realizations of each CMD drawn from its best-fit model, as
described in \cite{match}. It is again important to note the
difference between this approach and using stars of a particular
spectral type to make 3D maps of the Milky Way \citep[as in
e.g.][]{juric08}. The CMD fitting will give a direct, high
signal-to-noise map of the stellar mass density, even if the stellar
populations are spatially varying. This approach also utilizes the
information for stars of different spectral types (colors,
luminosities) simultaneously.

A comparison of the best-fit model with the CMD data for the region at
$l$=94\degr~and $b$=+30\degr~(the same field as shown the lower left
panel in Figure \ref{fig:examplecmds}) is presented in Figure
\ref{fig:examplehess}. Although the model is not perfect (which
perhaps should not be expected due to the simple model population
make-up), the general features of the density distribution are
well-reproduced.  The mean residuals are 25\% of the mean star counts
across the portion of the CMD used in the fit.

\section{Results}
\label{sec:results}

\begin{figure*}
\epsscale{0.95}
\plotone{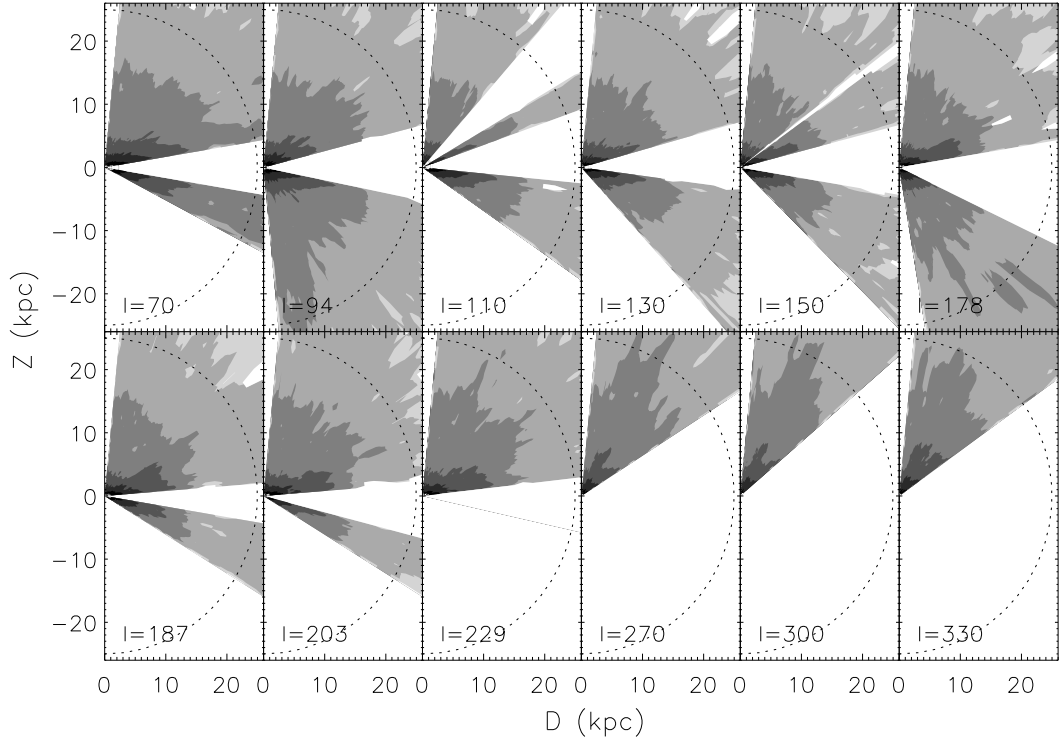}
\caption{ Tomography of the Milky Way. Shown are maps of the stellar
mass density in a set of slices of constant $l$, drawing on the single
population fits, as function of distance from the Sun, and height
above or below the Galactic plane, in kpc. Contours correspond to
steps of a factor 10 in stellar mass density. Each panel shows a
different imaging scan, the Galactic longitude of which is listed at
the bottom, and darker gray levels correspond to higher densities. The
dashed semicircles show the maximal distance to which fit results are
(mainly) based on main-sequence turn-off star colors and densities.}
\label{fig:densities}
\end{figure*}

\begin{figure*}
\epsscale{0.95}
\plotone{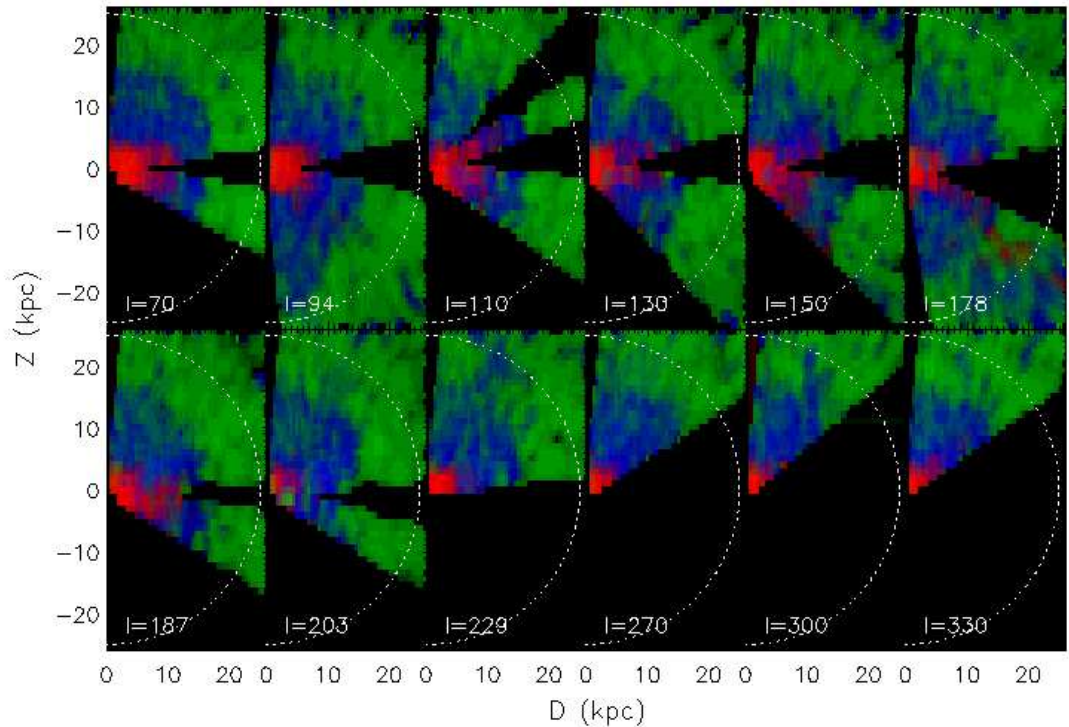}
\caption{ Density-metallicity tomography: shown is the distribution of
stellar populations from the triple population fits with the same
layout as in Fig. \ref{fig:densities}. The thick-disk-like population
with [Fe/H]$\simeq-$0.7 is color-coded as red, the inner halo-like
population with [Fe/H]$\simeq-$1.3 as blue, and the outer halo-like
population with [Fe/H]$\simeq-$2.2 as green. The white, dashed
semicircles show the distance up to where fit results are (mainly)
based on main-sequence turn-off star colors and densities.}
\label{fig:populations}
\end{figure*}

Rather than the stellar mass or the number of stars, the CMD fitting
software provides the star formation rate (SFR) in M$_\sun$yr$^{-1}$
corresponding to each population template for each distance modulus
bin. However, these SFRs can easily be converted to the spatial stellar mass
density using
\begin{equation}
\rho_{*,p} ~=~ \frac{3}{\omega (D_2^3 - D_1^3)} \times SFR_p~ \Delta t,
\end{equation}
where $SFR_p$ is the star formation rate in
$M_\sun~ yr^{-1}$ of template population $p$, $\Delta t$ is the
width of the age bin, $\omega$ is the solid angle in steradian
corresponding to each CMD, and $D_1$ and $D_2$ are the distances
corresponding to the limits of the distance modulus bin. The 2001
MATCH version uses a single power-law initial mass function
(IMF). In this analysis we opt for a Salpeter IMF
\citep{salpeter55} with cutoffs at 0.1 and 120 M$_\sun$, which yields
a total mass of about a factor 2 higher than currently favored IMFs
\citep[e.g.][]{kroupa93} for old populations \citep[see
also][]{MLfits}. 
It should also be kept in mind that these masses correspond to the
total mass of stars formed, which, especially for old populations,
does not correspond to the total mass in stars today, but rather to
the total mass in both stars and stellar remnants.

Figure \ref{fig:densities} shows the stellar mass densities obtained
from the CMD fits for each of the stripes, summed over all three
metallicities. Figure \ref{fig:populations} shows the same, but with a
color-coding that represents the relative contributions of the three
different template populations. Generally speaking, the stellar
density decreases in a continuous fashion with increasing distance
from the Galactic center and plane, although some deviations are
noticeable. For example, a region of apparent excess density extends
to the bottom of the $l$=94\degr~panel in Figure \ref{fig:densities},
caused by the Sagittarius stream \citep{majewski03,yanny09}.  The
contributions of the three different template populations change as
expected. With increasing distance from the Sun, the `thick disk',
`inner halo' and `outer halo' populations dominate in turn. The
Galactocentric distance where the metal-poor `outer halo' starts
dominating over the more metal-rich inner halo varies between 10 and
20 kpc.

\begin{figure*}
\epsscale{1}
\plotone{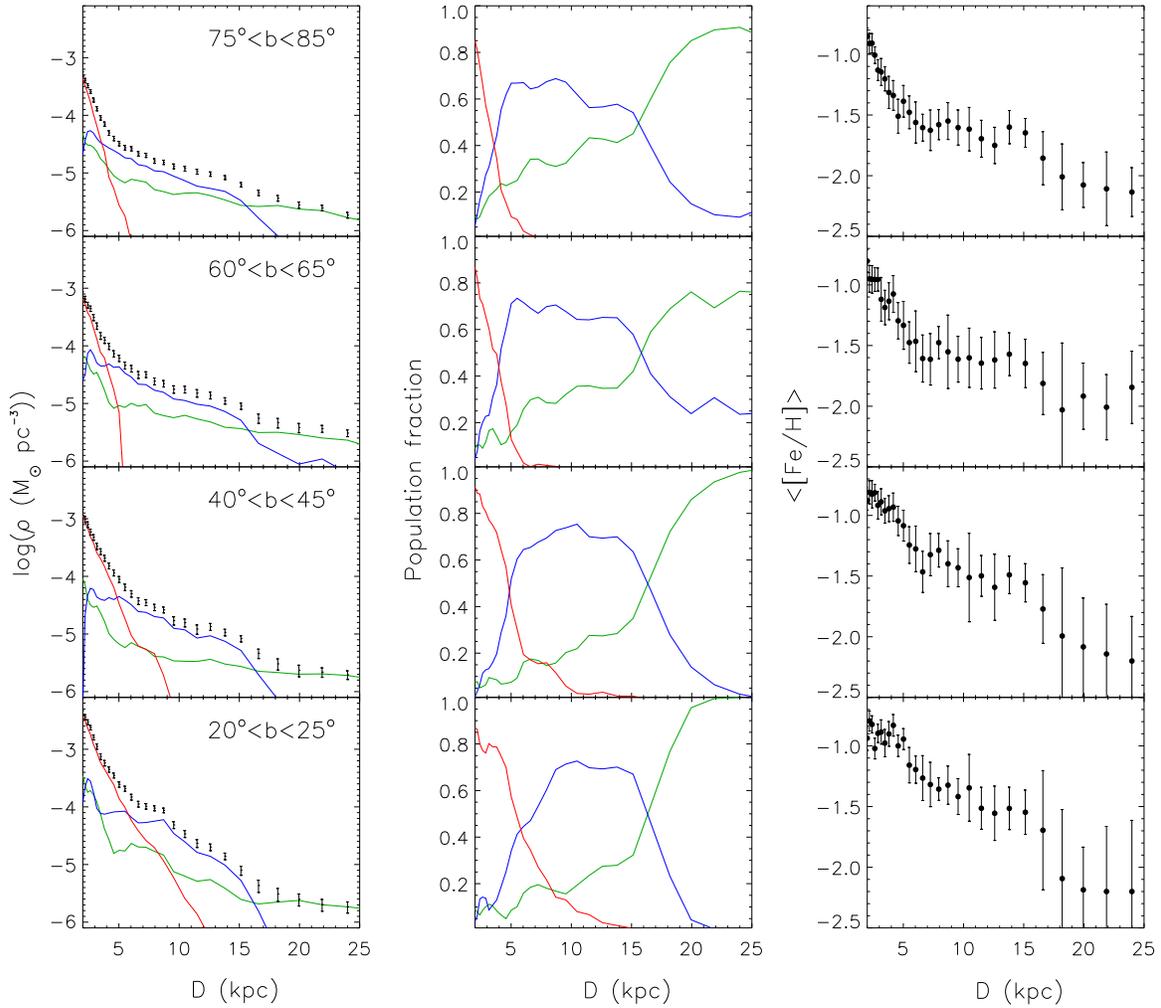}
\caption{ Mean stellar mass densities and metallicities as function of
distance. The left-hand panels show the total stellar mass density
with the black error bars for four different bins in Galactic
latitude, indicated in the top right of each panel, and averaged over
all longitudes. Colored lines show the mass density in the individual
template populations with red for the `thick-disk-like', blue for the
`inner-halo-like' and green for the `outer-halo-like' population.  In
the middle panels the colored lines indicate the fractional
contribution of the individual template populations to the total mass
density on a scale of 0 to 1, following the same color scheme as in
the left-hand panels.  The right-hand panels show the mass-weighted
mean metallicity.}
\label{fig:densitygraphs}
\end{figure*}

Figure \ref{fig:densitygraphs} shows the mean density and
mass-weighted metallicity as a function of distance from the Sun for a
set of different bins in Galactic latitude.  Apart from the total mass
density, the contributions from the individual template populations
are also plotted.  The increasing contribution of the thick disk to
the density along lines of sight with decreasing latitude is clearly
visible, both in the density graphs and in the steepness of the
metallicity gradient at distances less than 10 kpc. The metallicity
graphs indicate that the stellar halo has an average metallicity of
[Fe/H]$\simeq -$1.6 at distances less than $\sim$15 kpc, confirming
the findings of \cite{ivezic08}, but then drops to [Fe/H]$\lesssim -$2
further out. Although the transition is smooth, and both `halo-like'
template populations contribute to the mass density at every distance,
such a drop is consistent with the presence of two distinct
populations, as predicted by \cite{carollo07}. Recall, however, that
the transitions between different templates can only be taken as
direct evidence of a metallicity gradient in the halo, if age
differences between halo stars play no role in their turn-off
color. For the thick disk and stellar halo this is a reasonable
assumption, as both are generally believed to be relatively old
($\gtrsim$10 Gyr), so that the isochrones are much less sensitive to
age differences than to metallicity differences.

The stellar density values, $\rho_{*,p}$, are also presented in Table
\ref{tab:densities}. As the table contains 25\,476 rows, only the
first five rows are printed, with the complete table available in
electronic form online.

\begin{deluxetable*}{ccccccccc}
\tablecaption{Stellar density maps}
\tablewidth{0pt}
\tablehead{ \colhead{$l$} & \colhead{$b$} & \colhead{m-M} &
  \colhead{$\rho_1$} & \colhead{$\sigma_{\rho 1}$} &
  \colhead{$\rho_2$} & \colhead{$\sigma_{\rho 2}$} &
  \colhead{$\rho_3$} & \colhead{$\sigma_{\rho 3}$}\\
(\degr) & (\degr) & (mag) & ($M_\sun$ kpc$^{-3}$) & ($M_\sun$
  kpc$^{-3}$) & ($M_\sun$ kpc$^{-3}$) & ($M_\sun$ kpc$^{-3}$) &
  ($M_\sun$ kpc$^{-3}$) & ($M_\sun$ kpc$^{-3}$) }
\startdata
110.0 & $-$34.5 & 11.1 & 0.0 & 4.76E5 & 0.0 & 5.00E3 & 1.28E7 & 5.12E6\\
110.0 & $-$34.5 & 11.3 & 0.0 & 1.16E5 & 0.0 & 6.86E5 & 3.14E6 & 3.01E6\\
110.0 & $-$34.5 & 11.5 & 0.0 & 1.51E5 & 0.0 & 2.75E5 & 5.74E6 & 1.87E6\\
110.0 & $-$34.5 & 11.7 & 1.11E5 & 1.34E5 & 0.0 & 7.00E4 & 0.0 & 1.28E6\\
110.0 & $-$34.5 & 11.9 & 2.36E5 & 2.00E5 & 0.0 & 6.49E4 & 5.1E5 & 6.46E5\\
\enddata
\tablecomments{Densities $\rho_1$, $\rho_2$ and $\rho_3$ correspond to
  the mass densities for the template populations with [Fe/H]=$-$2.2,
  $-$1.3 and $-$0.7, respectively. The densities in this table
  correspond to the Salpeter IMF \citep{salpeter55} used by the MATCH
  software and are in solar masses per cubic kiloparsec, unlike the
  densities quoted in the text of the paper, which are corrected for a
  Kroupa IMF \citep{kroupa93} and in solar masses per cubic
  parsec. This table is available in its entirety in a
  machine-readable form in the online journal. A portion is shown here
  for guidance regarding its form and content}
\label{tab:densities}
\end{deluxetable*}

\subsection{Smooth Galactic model fits}
\label{sec:modelfits}

To describe the total stellar densities (i.e. summed over all three
template populations) quantitatively, we fit a model to all data
values $\rho_{*,i}$ represented by the maps in Figure
\ref{fig:densities}. Any substructures that are present on top of a
`smooth' stellar distribution should become apparent after subtraction
of the best-fit model. The model we use consists of a double
exponential thin disk and thick disk \citep{bahcall80,gilmore83}, and
an axisymmetric power-law halo \citep{ELS,chiba00,juric08}:
\begin{eqnarray}
\label{eq:galacticmodel}
\rho(R,Z) ~=~ \rho_{thin,\sun} \Bigg( e^{R_\sun/L_1}
exp\Big(-\frac{R}{L_1}-\frac{Z}{H_1}\Big) \nonumber \\
~+~ f_{thick,\sun}
e^{R_\sun/L_2} exp\Big(-\frac{R}{L_2}-\frac{Z}{H_2}\Big) \\
~+~ f_{halo,\sun} \Big(\frac{R_\sun}{\sqrt{R^2+(Z/q_h)^2}}\Big)^{n_h} \Bigg).
\nonumber
\end{eqnarray}
Here, $R_\sun$ is the distance from the Sun to the Galactic center;
$L_{thin}$, $H_{thin}$, $L_{thick}$ and $H_{thick}$ are the scale
lengths and heights of the thin and thick disk; $q_{halo}$ and
$n_{halo}$ are the halo flattening and power-law exponent;
$\rho_{thin,\sun}$ is the local thin-disk density and $f_{thick,\sun}$
and $f_{halo,\sun}$ are the local density fraction of the thick disk
and halo relative to the thin disk, respectively. In light of the
results of \cite{bell08} there seems to be no need for a more
sophisticated, for example triaxial, halo model. For a grid of
parameter values, models are compared to the density values resulting
from the CMD fits. Assuming $R_\sun$=7.6 kpc
\citep{vallee08}\footnote{As R$_\sun$ is still not very well
determined \citep[e.g.][]{bovy09}, we have also used values of 7.1 and
8.1 kpc for $R_\sun$ and found the effects of these changes on the fit
results to be negligible compared to the uncertainties quoted in Table
\ref{tab:modelfits}.}, the distance and the Galactic latitude and
longitude of each bin can be converted to $R$ and $Z$. Since the
density in each bin, $\rho_{*,i}(R,Z)$ has an uncertainty from the
Monte Carlo tests, the $\chi^2$ goodness-of-fit of each model can be
determined straightforwardly as 
\begin{equation}
\chi^2 ~=~ \sum_i \Big(\frac{\rho_{i,model}-\rho_{i,observed}}{\sigma_{i,observed}}\Big)^2.
\end{equation}
As mentioned before, we only use the bins for which 1.5$< D <$25 kpc
and $Z>$1.0 kpc. 

Some localized stellar overdensities are known to exist in the Milky
Way which may be strong enough to affect significantly the attempt to
fit a `smooth' model. The strongest one in the halo is the Sagittarius
stream, presumed to be tidal debris from the Sagittarius dwarf galaxy.
It arcs through the North Galactic cap through the stripes at
$l$=229\degr, 270\degr, 300\degr, and 330\degr at high latitudes
\citep{fieldofstreams}, and intersects the stripes at $l$=94\degr~and
178\degr~at negative latitudes \citep{majewski03,yanny09}. This can
already be seen in the density maps in Figure \ref{fig:densities}, for
example at $l$=94\degr, $Z<-$15 kpc and $D<$10 kpc. At lower latitudes
the largest known overdensity is the Monoceros overdensity, discovered
by \cite{newberg02} and of still controversial origin. It may
plausibly be part of the Low Latitude or anti-center stream, a
ring-like structure thought to encircle the Milky Way completely
\citep{ibata03}.  This structure can also be seen in Figure
\ref{fig:densities}, near the position of the original discovery (in
particular for the $l$=178\degr~and $l$=187\degr~stripes, where a
`bump' is apparent in the stellar density at 5$<D<$10 kpc and 0$<Z<$5
kpc).

To test for the influence of these structures on our fit results, we
fit the density maps twice, once with all data, and once without the
regions dominated by these structures. The regions that are excised to
remove the impact of the Sagittarius stream and Monoceros are listed
in Table \ref{tab:excised}, amounting to $\sim$4\% of all data points.

\begin{figure}[t]
\epsscale{1.1}
\plotone{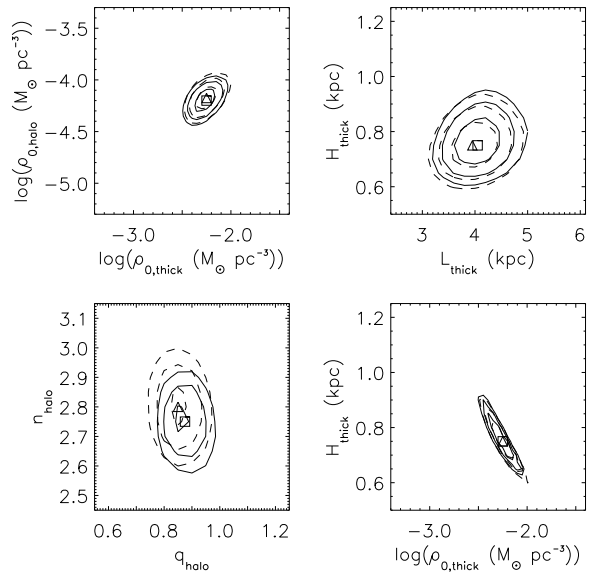}
\caption{
Results of fitting smooth Galactic models to the stellar density
maps. For the case where all data are 
used, the best fit values are indicated with squares and
iso-$\chi^2$ contours are plotted with solid lines. Triangles and
dashed contours are for the case when the Sagittarius stream and the
Monoceros overdensity are excised. In both cases, contours are scaled
to correspond to the 1, 2 and 3$\sigma$ uncertainties determined
through bootstrap tests.
}
\label{fig:modelfits}
\end{figure}

\begin{figure}[t]
\epsscale{1.0}
\plotone{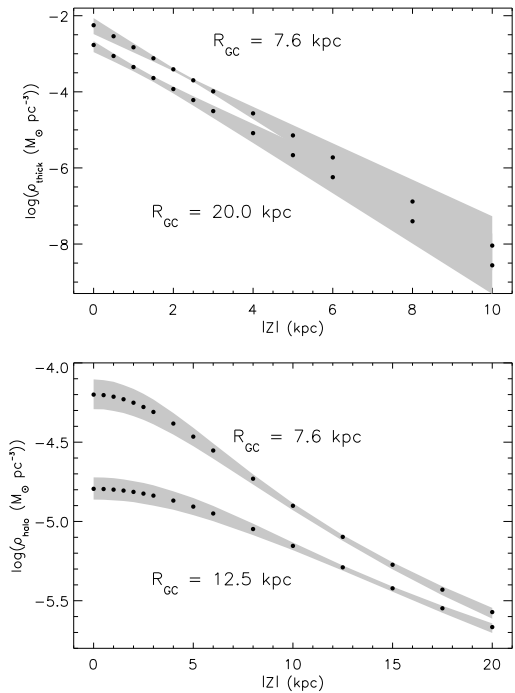}
\caption{ Thick disk (top) and halo (bottom) densities in the best-fit
smooth model as function of distance from the Galactic plane. The
densities were calculated for two galactocentric radii, 7.6 kpc (the
assumed galactocentric radius of the Sun) and 12.5 kpc, and the
sequences of points are labeled accordingly. Dots show the density at
each location for the best-fit model, while the grey areas show the
95\% confidence region determined from the bootstrap analysis.  }
\label{fig:denserrors}
\end{figure}

The role of the thin disk is limited by only looking at the area at
$|Z|>$1.0 kpc, and we fix its parameters at $L_{thin}$=2.6 kpc and
$H_{thin}$=0.25 kpc, following \cite{juric08} and in line with earlier
star count results \citep[e.g][]{siegel02}. In other words, we treat
the thin disk as a known and fixed `perturbation' to our CMD
analysis. We are now left with seven free parameters in fitting the
total stellar mass density (summed over the three metallicity
components): the thick-disk parameters $f_{thick}$, $L_{thick}$ and
$H_{thick}$, the halo parameters $f_{halo}$, $q_{halo}$ and
$n_{halo}$, and the local thin-disk density $\rho_{thin,\sun}$. The
latter serves as normalization for the thick disk and halo density,
hence in practice reducing the number of free parameters to six.

\subsection{Structural parameters of the thick disk and stellar halo}
\label{sec:structpar}

In Table \ref{tab:modelfits} we list the parameter values
corresponding to the best fits, with the mass densities scaled to a
Kroupa-like IMF \citep{kroupa93}. The best fit to all data has a
reduced $\chi^2$ of 4.2, while the best fit to the data without
Sagittarius and Monoceros has a reduced $\chi^2$ of 3.9. Inclusion of
these structures thus has a noticeable impact on the goodness-of-fit,
but in both cases the smooth models provide a poor description of the
data, in line with \cite{bell08} and \cite{juric08}.  Since this
inhibits deriving proper confidence intervals based on the $\chi^2$
statistic, the uncertainties on the fit parameters listed in Table
\ref{tab:modelfits} were obtained by resampling the data points,
$\rho_{*,i}(R,Z)$, as in bootstrap tests.  For each test, resampled
data sets were created by randomly drawing data points (with
replacement) from our set of stellar densities, $\rho_{*,i}(R,Z)$,
until the same number of data points is reached. The fit parameters
were then re-determined for the resampled data sets by re-fitting the
model given by Equation \ref{eq:galacticmodel}. This procedure was
repeated 200 times for the full data set and for the data set without
Sagittarius and Monoceros. In Figure \ref{fig:modelfits} we present
the contour plots of the 1$\sigma$, 2$\sigma$, and 3$\sigma$ (68\%,
95\% and 99.7\%) confidence levels around the fit parameters, which
are also based on the bootstrap analysis.

The mass densities of the thick disk and halo are very well
constrained, as demonstrated in Figure \ref{fig:denserrors}, where the
best model densities and their 95\% confidence intervals are plotted
at different locations. The thick-disk density is best constrained at
($R_{GC}$,$|Z|$)=(7.6,2.5) kpc as $\rho_{thick}=10^{-3.70\pm0.02}$
$M_\sun$pc$^{-3}$, and the halo density at
($R_{GC}$,$|Z|$)=(10.0,12.5) kpc as $\rho_{halo}=10^{-5.19\pm0.01}$
$M_\sun$pc$^{-3}$ ($10^{-5.21\pm0.01}$ $M_\sun$pc$^{-3}$ when
Sagittarius and Monoceros are omitted). These formal errors are
presumably exceeded by uncertainties in the IMF and associated with
substructure. At the solar location the uncertainties on the density
are significantly larger than at the locations stated above, as the
local values are not directly constrained by the data. Rather, they
are extrapolations from data at higher $|Z|$ and $R_{GC}$, and are
therefore model dependent. However, since most previous work has
focussed on local samples of stars, we will use the local
extrapolations of our results in order to compare with earlier
studies.  The value we find for the local mass density of the thick
disk is $\rho_{thick,\sun}=10^{-2.3\pm0.1}$ $M_\sun$pc$^{-3}$, and for
the stellar halo $\rho_{halo,\sun}=10^{-4.20\pm0.05}$
$M_\sun$pc$^{-3}$ ($\rho_{thick,\sun}=10^{-2.3\pm0.1}$
$M_\sun$pc$^{-3}$ and $\rho_{halo,\sun}=10^{-4.18\pm0.05}$
$M_\sun$pc$^{-3}$ when Sagittarius and Monoceros are omitted).

Recent dynamical modeling \citep[e.g.][]{flynn06} implies a local
thick-disk density of $10^{-2.46} M_\sun$pc$^{-3}$ and a local halo
density of $10^{-4} M_\sun$pc$^{-3}$, similar to our
findings. However, our local mass densities include stellar remnants,
and a proper comparison should take this into account.  The total
local mass density, dominated by the thin disk (in stars and white
dwarfs), is $\sim0.038 M_\sun$pc$^{-3}$ \citep{flynn06}.  Star-count
studies typically yield thick-disk and halo normalizations with
respect to the thin disk of 1 to 10\% and 0.1 to 0.2\%, respectively
\citep{siegel02}. This therefore implies a local density, including
remnants, of $10^{-3.40}$ to $10^{-2.42}$ $M_\sun$pc$^{-3}$ for the
thick disk and $10^{-4.42}$ to $10^{-4.12}$ $M_\sun$pc$^{-3}$ for the
stellar halo, values that are consistent with our density
estimates. If we normalize our findings by the local value from
\cite{flynn06}, we find local thick-disk and halo mass-density
fractions of 15$\pm$4\% and 0.17$\pm$0.03\%, respectively.
\cite{juric08}, in their analysis of SDSS data, find 12$\pm$1\% for
the thick disk, consistent with our results, and 0.51$\pm$0.13\% for
the halo, inconsistent at the 2.6$\sigma$ level.  These normalizations
also imply that the stellar density of the thin disk should never
amount to more than $\sim$10\% at $|Z| > $1.5 kpc in our maps,
confirming that it is merely a perturbation to our analysis at these
locations.

Thick-disk structural parameters determined previously through
star-count studies have shown a large spread \citep[e.g.][]{siegel02}.
Scale heights vary from $\sim$0.6 to $\sim$2.0 kpc, with the more
recent measurements converging to the lower end of this range; and
scale lengths vary from 2.8 to 4.3 kpc. \cite{juric08} find very
different values depending on the tracer stars or photometric parallax
relations that are used, but settle on 0.9$\pm$0.2 kpc and 3.6$\pm$0.7
kpc for the thick-disk scale height and length. Through our present
analysis we have been able to estimate the scale height and length of
the thick disk to 10\% accuracy: $H_{thick}=0.75\pm0.07$ kpc and
$L_{thick}=4.1\pm0.4$ kpc. This is consistent with but more precise
than previous determinations, and in good agreement with the work of
\cite{juric08}.

Halo parameters, when described by a single power-law, are probably
varying with radius \cite[e.g.][]{bell08,watkins09,sesar10}, and in
this study only the stellar halo within a radius of 25 kpc from the
Sun has been considered. Two recent studies based on SDSS data have
measured the flattening and power-law index of the stellar
halo. \cite{juric08}, reaching to $\simeq$20 kpc, find a halo
flattening parameter of $q_h$=0.64 and a power-law index of $n_h$=$-$2.77
with quoted uncertainties of $\lesssim$0.1 and $\lesssim$0.2,
respectively. \cite{bell08} find that a $q_h$ of 0.6 to 0.7 gives the
least excess rms scatter around their model fits, but that a $q_h$ of
0.7 to 0.8 gives the best $\chi^2$ values. In either case, the
constraints on $q_h$ are weak. The halo model used by \cite{bell08}
is a broken power-law, with independent indices on either side of the
break radius. These parameters are also weakly constrained, with the
preferred break radius lying between 20 and 30 kpc, the inner
power-law index between 2 and 3, and the outer power-law index between
3 and 4. Similarly, \cite{watkins09} use RR Lyrae variable stars in
SDSS Stripe 82 to infer $n_h$=$-$2.4 within 23 kpc and $n_h$=$-$4.5 at
larger distances. By comparison, our analysis yields very precise
measurements of $n_{h}=-2.75\pm0.07$ and $q_{h}=0.88\pm0.03$ for the
stellar halo within 25 kpc from the Sun. While our value for the
power-law index is in excellent agreement with \cite{juric08} and the
inner power-law index from \cite{bell08}, we find the stellar halo to
be less flattened. Note that these previous SDSS determinations were
restricted to $b\gtrsim$30\degr; the inclusion of the SEGUE scans
across the Galactic plane boosts the power to determine flattening.

\begin{deluxetable}{lccc}
\tablecaption{Regions Excised for Second Model Fit}
\tablewidth{0pt} 
\tablehead{ \colhead{Structure} & \colhead{$l$ (\degr)} & \colhead{$b$ (\degr)} & \colhead{$m-M$ (mag)}}
\startdata
Sgr stream & 94 & $<-$65 & $>$16.0 \\
Sgr stream & 178 & $-$65 to $-$25 & $>$16.5 \\
Sgr stream & 229 & +55 to +70 & $>$15.5 \\
Sgr stream & 270 & +55 to +70 & $>$15.5 \\
Sgr stream & 300 & +58 to +72 & $>$15.5 \\
Sgr stream & 330 & +60 to +80 & $>$15.5 \\
Monoceros & 178 & +15 to +35 & 14.0--16.0 \\
Monoceros & 187 & +15 to +35 & 14.0--16.0 \\
\enddata
\label{tab:excised}
\end{deluxetable}

\begin{deluxetable}{lcc}
\tablecaption{Smooth Galactic model fits}
\tablewidth{0pt} 
\tablehead{ \colhead{Parameter} & \colhead{Full Data} &
  \colhead{Exclude Sgr,Mon}}
\startdata
$\rho_{thick,\sun}$ ($M_\sun$pc$^{-3}$) & $10^{-2.3\pm0.1}$ & $10^{-2.3\pm0.1}$ \\
$L_{thick}$ (kpc) & 4.1$\pm$0.4 & 4.0$\pm$0.4 \\
$H_{thick}$ (kpc) & 0.75$\pm$0.07 &  0.75$\pm$0.08 \\
$\rho_{halo,\sun}$ ($M_\sun$pc$^{-3}$) & $10^{-4.20\pm0.05}$ & $10^{-4.18\pm0.05}$ \\
$q_{halo}$ & 0.88$\pm$0.04 & 0.85$\pm$0.03 \\
$n_{halo}$ & $-$2.75$\pm$0.07 & $-$2.80$\pm$0.07 \\
\enddata
\label{tab:modelfits}
\end{deluxetable}

\subsection{Substructure, or deviations from a smooth model}
\label{sec:substructure}

\begin{figure*}
\epsscale{1.0}
\plotone{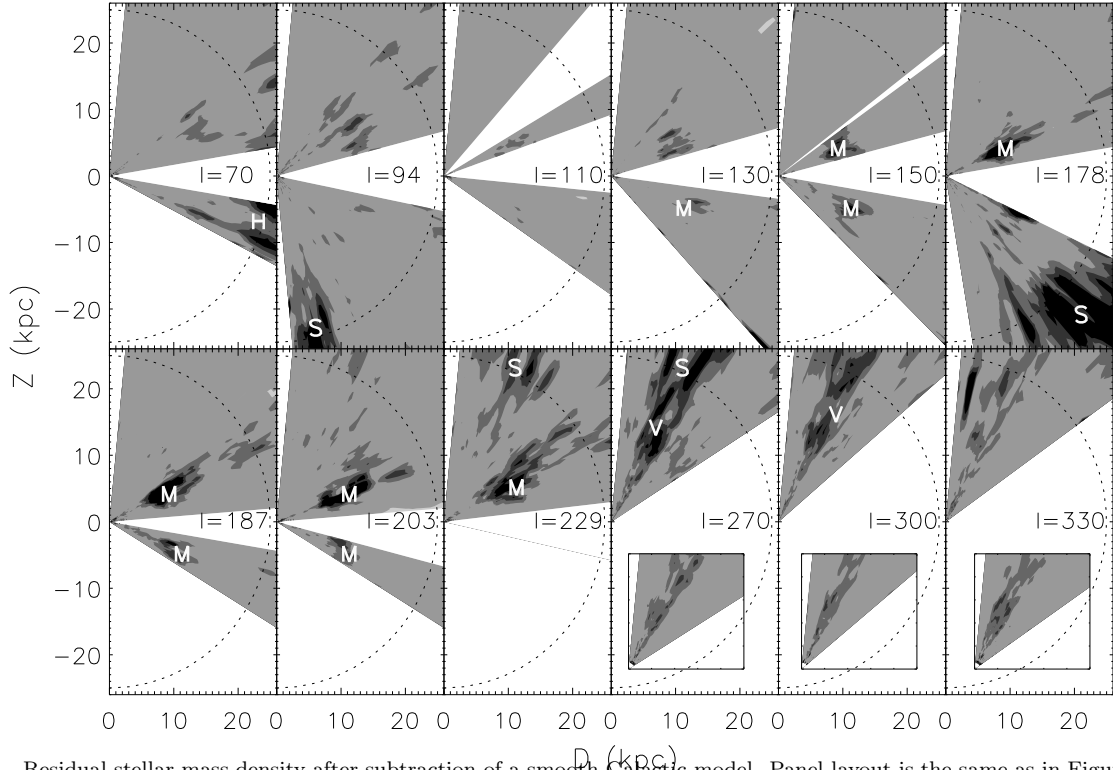}
\caption{ Residual stellar mass density after subtraction of a smooth
Galactic model. Panel layout is the same as in Figure
\ref{fig:densities}. Starting from black, the grayscale levels
correspond to areas with residual densities $>3$, $3$--$2$, $2$--$1$,
$1$ to $-1$, and $<-1$ times the model density. White areas contain no
information due to an absence of data or too high reddening. The
dashed semicircles show the distance up to where fit results are
(mainly) based on main-sequence turn-off star colors and
densities. Four known overdensities are labeled: the Sagittarius
stream (S), the Virgo overdensity (V), the Hercules-Aquila cloud (H),
and the Monoceros, or Low Latitude stream (M). The panels for
$l$=270\degr, 300\degr and 330\degr~have insets showing the inner
5$\times$5 kpc region. }
\label{fig:subtracted}
\end{figure*}

To focus on substructure in the stellar halo and thick disk we now
subtract the best-fit smooth model to the complete data set from the
density maps in Figure \ref{fig:densities}.  The residuals after
subtracting this smooth model are shown in Figure
\ref{fig:subtracted}, revealing a wealth of substructure. In
particular, many overdense regions jump out, and although there are
also underdense regions, the density contrast of the latter with
respect to the smooth model appears to be much weaker.  In principle
the CMD fitting method can provide not only the stellar densities, but
also yield information on the stellar populations of the
overdensities, for example their metallicities. However, the current
analysis uses three simple template populations, which differ in
metallicity but all have the same age ($\sim$14 Gyr). For the thick
disk and stellar halo it is justifiable to interpret the population
differences detected in our fits as metallicity differences, as both
components consist of old stars. Overdensities, however, might be
streams populated by extra-galactic sources or material swept up from
even the thin disk. There is no reason to believe that their stars
should have the same ages as the smooth components on which they are
superposed. The reconstruction of the detailed stellar population
properties of specific substructures thus requires a more detailed
analysis of each overdensity. Therefore, we limit ourselves here to
listing the overdensities we detect and a short description of the
main structures responsible for the overdensities seen in Figure
\ref{fig:subtracted}.

As the strength of an overdensity detection depends on the smooth
model that is subtracted from the density maps and each model
parameter has a specific uncertainty, we need to define what we
consider a significant detection. We do this by constructing a set of
smooth Galactic models on a grid of parameters, and selecting all
models which are within 1$\sigma$ of the best-fit model in the 6D
parameter space given by the model parameters in Table
\ref{tab:modelfits}.  Density maps are evaluated for all pixels in
these 282 models.  The pixels with densities at least 3$\sigma$ higher
than the model density for more than half (141) of the models are
considered to be significant.  Table \ref{tab:overdensities} lists all
significantly detected overdensities, most of which consist of a large
number of such pixels.  The peak significance listed in the table
corresponds to the most significant pixel in each overdensity with
respect to the best-fit smooth model.  The total significance of the
overdensity is much larger in most cases, as they consist of many
pixels. Because the metallicities and ages of the stars in the
overdensities do not necessarily fall neatly within one of the
template populations used in the CMD fits, the distances listed should
be taken as estimates, not as precise measurements. Note that one
detection is caused by two bright globular clusters, NGC 5024 (M53)
and NGC 5053, which happen to lie exactly in the scan at $l$=330\degr.

{\it Sagittarius stream.}  Probably the most striking stellar
`substructure' in the Galactic halo, the Sagittarius stream (labeled
with an `S' in Fig. \ref{fig:subtracted}) is visible in several
stripes, both in the northern and the southern Galactic hemisphere.
The northern arm is clearly visible in the stripes at $l$=229\degr,
270\degr, and 300\degr, just outside the dashed semicircles
marking a distance of 25 kpc. At $l$=330\degr~the stream is more
distant \citep{fieldofstreams} and off the plot. The two arms in
which this part of the stream is split \citep{fieldofstreams} are
detected very clearly, particularly in the $l$=270\degr
stripe. Typical peak densities in these stripes are 5 to
8$\times10^{-6} M_\sun$pc$^{-3}$. At $l$=203\degr~the northern Sgr
stream should enter the stripe around $b$=30\degr~and plunge into the
disk \citep{fieldofstreams,yanny09}. As it is roughly
equidistant with the Low-Latitude Stream, it is difficult to
distinguish it in the overdensity at low latitudes in this stripe.  In
the southern Galactic cap, the Sgr stream yields strong detections in
two stripes, namely the $l$=94\degr~and 178\degr~stripes, reproducing
the detection by \cite{yanny09}. Here the stream is also seen at
distances of 20 kpc and higher. The density of the stream in these
southern detections peaks at $\sim$1$\times10^{-5} M_\sun$pc$^{-3}$,
slightly higher than in the North Galactic cap.

{\it Virgo.}  The Virgo overdensity \citep{juric08} is another strong
overdensity in the Northern Galactic cap, centered at
($l$,$b$)$\simeq$(280\degr,70\degr) and extending between distances of
$\sim$10 to $\sim$20 kpc \citep{bell08,juric08}. It is very prominent
in the residual maps of the stripes at $l$=270\degr~and 300\degr~in
the same direction as the Sgr stream, but at distances less than 20
kpc. The overdensities at similar distances in the $l$=330\degr~stripe
might be related to Virgo as well. Typical densities in the Virgo
overdensity we find in the $l$=270\degr~and 300\degr~stripes are
$\sim$2$\times10^{-5} M_\sun$pc$^{-3}$, peaking at around
$\sim$4$\times10^{-5} M_\sun$pc$^{-3}$.

{\it Hercules-Aquila.}  A large overdensity extending both above and
below the disk, and spanning $\sim$80\degr~in longitude, was detected by
\cite{hercaqui} and christened the Hercules-Aquila cloud. This
structure is centered at $l \simeq$40\degr~, but is visible out to $l
\lesssim$90\degr~\citep{hercaqui}. The overdensities detected
in the stripe at $l$=70\degr~ and distances larger than $\sim$22 kpc
might therefore be associated with this recently discovered stellar
component.

{\it Monoceros and Low Latitude Stream.}  The Monoceros structure was
first discovered in SDSS data by \cite{newberg02} as a strong stellar
overdensity at ($l$,$b$)$\simeq$(200\degr,20\degr).  Later such
overdensities were detected over a large range of Galactic longitudes
both above and below the Galactic plane, always at low latitudes and
roughly equidistant, thus forming a ring-like structure around the
Milky Way \citep{ibata03,yanny03,conn05,conn07}. In the stripes at the
longitudes close to the original discovery, $l$=178\degr, 187\degr,
203\degr and 229\degr, the overdensity is detected at high
significance at heliocentric distances of roughly 10 kpc and $\sim$3
kpc above the Galactic plane. The ring-like nature of the Low Latitude
Stream appears to be confirmed by our new map, as overdensities at the
same heliocentric distances and latitudes are present in the stripes
at $l$=94\degr, 130\degr~and 150\degr, but not at $l$=110\degr. The
density of the structure is highest in the stripes at $l$=178\degr,
187\degr, and 203\degr, where the maximum density reached is
$\sim$1.5$\times10^{-4} M_\sun$pc$^{-3}$.  The density appears to fall
off by a factor of $\sim$2 in the stripes at both lower and higher
longitudes.  The overdensity detections at negative latitudes
\citep{ibata03,conn05,conn07} are confirmed here, with an overdensity
detected consistently $\sim$4 kpc below the plane from $l$=130\degr~to
$l$=203\degr. Below the plane the overdensities have lower densities
than above the plane, increasing from $\sim$0.2$\times10^{-4}
M_\sun$pc$^{-3}$ at $l$=130\degr~ to $\sim$1.3$\times10^{-4}
M_\sun$pc$^{-3}$ at $l$=130\degr.

{\it Detection at (R,Z)=(6.5,1.5).} \cite{juric08} detected an
overdensity at Galactocentric coordinates (R,Z)=(6.5,1.5) kpc. In the
stripes at $l$=270\degr, 300\degr~and 330\degr~we find a very nearby
overdensity at $b\simeq$65\degr~and at a distance range from 4 kpc to
the smallest distance we probe, 1.5 kpc. This overdensity therefore
corresponds to the same as detected by \cite{juric08}. The nature of
this overdensity is so far unknown.

{\it Unknown substructure.}  In a few locations less prominent
substructures are visible, that cannot directly be identified with
known overdensities. Most of these are located at low latitudes and
might be related to the Low Latitude Stream. In a future publication
we will study these overdensities and their possible connection with
the Low Latitude Stream in detail.

\begin{deluxetable}{ccccc}
\tablecaption{Detected overdensities}
\tablewidth{0pt} 
\tablehead{ \colhead{$l$ (\degr)} & \colhead{$b$ (\degr)} &
  \colhead{$D$ (kpc)} & \colhead{$\sigma_{peak}$} & \colhead{ID} }
\startdata
70  & ($-$23,$-$21)  & (14,16)  & 5.7  &  \\
70  & ($+$36,$+$38)  & (15)     & 8.5  &  \\
70  & ($-$23,$-$22)  & (22,25]  & 4.8  & Hercules-Aquila? \\
70  & ($-$13,$-$10]  & (23,25]  & 3.4  & Hercules-Aquila? \\
70  & [$+$10,$+$12)  & (23,25]  & 3.5  & Hercules-Aquila? \\
94  & ($-$82,$-$70)  & (21,25]  & 6.7  & Sgr stream \\
94  & ($+$18,$+$22)  & (12,14)  & 4.1  &  \\
94  & ($+$32,$+$35)  & (14,16)  & 8.4  &  \\
130 & ($-$18,$-$16)  & (13)     & 4.2  &  \\
130 & ($+$19,$+$22)  & (8,13)   & 4.7  &  \\
130 & ($+$26,$+$27)  & (9)      & 7.1  &  \\
130 & ($+$35,$+$37)  & (9,13)   & 5.0  &  \\
150 & ($-$31,$-$21)  & (11,14)  & 3.9  &  \\
150 & ($+$17,$+$29)  & (8,13)   & 140  & Monoceros \\
178 & ($-$62,$-$39)  & (22,25]  & 6.2  & Sgr stream \\
178 & ($-$59.5)      & (14)     & 3.1  &  \\
178 & ($-$48,$-$46)  & (11.5)   & 3.5  &  \\
178 & ($-$42,$-$40)  & (12.5)   & 3.2  &  \\
178 & ($-$30,$-$26]  & (11,15)  & 8.3  &  \\
178 & ($+$16,$+$32)  & (5,14)   & 9.7  & Monoceros \\
187 & ($-$30,$-$17)  & (8,13)   & 5.7  &  \\
187 & ($+$14,$+$35)  & (6,15)   & 13.1 & Monoceros \\
203 & ($-$28,$-$15]  & (8.5,13) & 8.5  & Monoceros \\
203 & ($+$14,$+$31)  & (7,15)   & 22   & Monoceros \\
203 & ($+$21.5)      & (20)     & 3.3  &  \\
229 & ($+$15,$+$40)  & (7,18)   & 7.1  & Monoceros \\
229 & ($+$35,$+$38)  & (21,24)  & 6.5  &  \\
229 & ($+$57,$+$64)  & (23,25]  & 4.2  & Sgr stream \\
270 & ($+$58,$+$69)  & (23,25]  & 4.1  & Sgr stream \\
270 & ($+$55,$+$74)  & (2,23)   & 8.8  & Virgo/(6.5,1.5) \\
300 & ($+$65,$+$70)  & [1.5,4)  & 5.4  & (6.5,1.5) \\
300 & ($+$58,$+$72)  & (10,19)  & 16.1 & Virgo \\
300 & ($+$60,$+$65)  & (21,25]  & 4.1  & Virgo/Sgr stream \\
330 & ($+$57,$+$72)  & [1.5,4)  & 5.5  & (6.5,1.5) \\
330 & ($+$55,$+$73)  & (12,24)  & 7.2  & Virgo \\
330 & ($+$78,$+$81)  & (16.5,22) & 12.2 & NGC5024/NGC5053 \\
\enddata
\tablecomments{For Galactic latitude $b$ and distance $D$ ranges are
  given within brackets, with square brackets indicating an edge of
  the data rather than the edge of a detection.}
\label{tab:overdensities}
\end{deluxetable}

\section{Summary and Conclusions}
\label{sec:discussion}

We have applied CMD-fitting techniques to SEGUE photometric data to
study the thick disk and stellar halo of the Galaxy at both high and
low latitudes. Three template stellar populations, all with an age of
$\sim$14 Gyr and metallicities of [Fe/H]=$-$0.7, $-$1.3 and $-$2.2,
are fit to the data, yielding a three-dimensional census of stellar
mass. Since the thick disk and stellar halo consist of predominantly
very old stars, differences in turn-off color are indicative of
differences in metallicity. A change of turn-off color is seen in the
halo at distances of 15 to 20 kpc, which is therefore interpreted as a
change in metallicity. At distances smaller than $\sim$10 kpc our
results indicate a mean metallicity of [Fe/H]$\sim-$1.6, similar to
the work by \cite{ivezic08}, who find a value of [Fe/H]$\sim-$1.5 out
to $\sim$9 kpc. At larger distances, D$\gtrsim$15 kpc, our results
indicate that the mean metallicity of the stellar halo is
[Fe/H]$\sim-$2.2, however. This transition is consistent with the
inference from local samples of stars (D$\lesssim$4 kpc) by
\cite{carollo07} and preliminary results from \cite{beers09}, but this
is the first time this change in population properties is measured
{\it in situ}.

Structural parameters of the Galaxy can be derived
from the resulting stellar-density maps through the comparison with
models. Our fits of models with an exponential thin and thick disk and
a power-law halo (Eq. \ref{eq:galacticmodel}) yield constraints on the
thick disk local density, $\rho_{thick,\sun}$, scale height,
$H_{thick}$, and scale length, $L_{thick}$, and the halo local
density, $\rho_{halo,\sun}$, flattening, $q_{halo}$ and power-law
index, $n_{halo}$.  As pointed out by \cite{fieldofstreams},
\cite{bell08}, and \cite{juric08}, the SDSS data of the stellar halo
contains clear evidence of the presence of substructure with respect
to smooth halo models. \cite{bell08} also demonstrated that the larger
known substructures in the halo significantly influence their model
fits. To test the sensitivity of our model fits to known substructure,
we perform our fits to the full data set, as well as to a data set
with the Sagittarius stream and Monoceros overdensity removed. The
results are tabulated in Table \ref{tab:modelfits}, and are not
strongly influenced by the presence of these structures. In either
case, the reduced $\chi^2$ of the best fit, 4.2 and 3.9, shows that
the smooth model is a poor fit to the data.

The values we find for the local density of the thick disk,
$\rho_{thick,\sun}=10^{-2.3\pm0.1}$ $M_\sun$pc$^{-3}$ and the local
halo density, $\rho_{halo,\sun}=10^{-4.20\pm0.05}$ $M_\sun$pc$^{-3}$,
agree well with star-count studies. In addition, our results for the
thick-disk structural parameters, $H_{thick}=0.75\pm0.07$ kpc and
$L_{thick}=4.1\pm0.4$, are well within the range of parameters found
in previous studies \citep{siegel02} and consistent with the results
of \cite{juric08}. For the flattening of the stellar halo, where we
find $q_{halo}=0.88\pm0.04$, our analysis seems to diverge from the
work by \cite{bell08} and \cite{juric08}, who use SDSS data to study
the halo out to similar distances. On the other hand, the power-law
index we recover, $n_{halo}=-2.75\pm0.07$, is in excellent agreement.

Subtracting the best model from the stellar-density maps unveils
abundant substructure (Fig. \ref{fig:subtracted}). Most of the
overdensities seen can be attributed to know structures in the halo
and outer disk of the Galaxy: the Sagittarius stream, Virgo
overdensity, and the Monoceros overdensity or Low Latitude Stream.
Whereas the origin of the Sagittarius stream is known to be the result
of the disruption of the Sagittarius dwarf galaxy, the nature of the
other two entities is currently under debate, in particular the
interpretation of the overdensities at low Galactic latitudes remains
contentious.  A detailed analysis of all substructures is outside the
scope of this paper, but an in-depth study of the low latitude
substructure is planned to be presented in a future publication.

\acknowledgments
The authors thank Eric Bell and Constance Rockosi for stimulating and
helpful conversations and Sergey Koposov for logistical help. We are
also grateful for helpful comments and feedback from \v{Z}eljko Ivezi\'c
and Mario Juri\'{c}.
J.T.A.d.J was supported by DFG Priority Program 1177.
T.C.B. acknowledges partial support for this work from grant 
PHY 08-22648: Physics Frontier Center/Joint Institute for Nuclear
Astrophysics (JINA), awarded by the U.S. National Science Foundation.

Funding for the SDSS and SDSS-II has been provided by the Alfred
P. Sloan Foundation, the Participating Institutions, the National
Science Foundation, the U.S. Department of Energy, the National
Aeronautics and Space Administration, the Japanese Monbukagakusho, the
Max Planck Society, and the Higher Education Funding Council for
England. The SDSS Web Site is http://www.sdss.org/.

The SDSS is managed by the Astrophysical Research Consortium for the
Participating Institutions. The Participating Institutions are the
American Museum of Natural History, Astrophysical Institute Potsdam,
University of Basel, University of Cambridge, Case Western Reserve
University, University of Chicago, Drexel University, Fermilab, the
Institute for Advanced Study, the Japan Participation Group, Johns
Hopkins University, the Joint Institute for Nuclear Astrophysics, the
Kavli Institute for Particle Astrophysics and Cosmology, the Korean
Scientist Group, the Chinese Academy of Sciences (LAMOST), Los Alamos
National Laboratory, the Max-Planck-Institute for Astronomy (MPIA),
the Max-Planck-Institute for Astrophysics (MPA), New Mexico State
University, Ohio State University, University of Pittsburgh,
University of Portsmouth, Princeton University, the United States
Naval Observatory, and the University of Washington.

\end{document}